\documentclass[twocolumn]{aastex63}

\usepackage{todonotes}
\usepackage{amsmath}

\newcommand{\bx}{\mathbf{x}}

\newcommand{\bxi}{\boldsymbol{\xi}}
\newcommand{\bk}{\mathbf{k}}
\newcommand{\bq}{\mathbf{q}}
\newcommand{\pone}{p$_1$}
\newcommand{\ptwo}{p$_2$}
\newcommand{\pthree}{p$_3$}
\newcommand{\Bcoeff}{$B^\sigma_{\bk\bq}$}

\received{\today}
\revised{}
\accepted{}
\submitjournal{APJ}

\shorttitle{Analyzing supergranular power spectra using helioseismic normal-mode coupling}
\shortauthors{Hanson, Hanasoge and Sreenivasan}

\begin{document}

\title{Analyzing supergranular power spectra using helioseismic normal-mode coupling}

\correspondingauthor{Chris S. Hanson}
\email{hanson@nyu.edu}

\author{Chris S. Hanson}
\affiliation{Center for Space Science, NYUAD Institute\\ New York University
Abu Dhabi\\ Abu Dhabi, UAE}

\author{Shravan Hanasoge}
\affiliation{Department of Astronomy and Astrophysics\\ Tata Institute of Fundamental Research\\Mumbai, India}
\affiliation{Center for Space Science, NYUAD Institute\\ New York University
Abu Dhabi\\ Abu Dhabi, UAE}

\author{Katepalli R. Sreenivasan}
\affiliation{Department of Physics, Courant Institute of Mathematical Sciences, Tandon School of Engineering\\ New York University, New York, 10003}
\affiliation{Center for Space Science, NYUAD Institute\\ New York University
Abu Dhabi\\ Abu Dhabi, UAE}



\begin{abstract}

Normal-mode coupling is a technique applied to probe the solar interior using surface observations of oscillations. {The technique, which is straightforward to implement, makes more use of the seismic information in the wavefield than other comparable local imaging techniques and therefore has the potential to significantly improve current capabilities.} Here, we examine supergranulation power spectra using mode-coupling analyses of intermediate-to-high-degree modes by invoking a Cartesian-geometric description of wave propagation under the assumption that the localized patches are much smaller in size than the solar radius. We extract the supergranular power spectrum and compare the results with prior helioseismic studies. {Measurements of the dispersion relation and life times of supergranulation, obtained using near surface modes (f and \pone{}), are in accord with the literature}. We show that the cross-coupling between the \ptwo{} and \pthree{} acoustic modes, which are capable of probing greater depths, are also sensitive to supergranulation.

\end{abstract}

\keywords{Methods: data analysis -- Sun: helioseismology -- convection -- waves}


\section{Introduction}
Helioseismology is applied to image the solar interior through the analysis of the Sun's observed surface wave field \citep[e.g.][]{jcd_review_2002}. The oscillations, which comprise acoustic p modes and surface-gravity f modes, are thought to be generated by near-surface turbulent stresses \citep{goldreich_keely_1977b}. Spherically symmetric models \citep[e.g.][]{jcd_etal_1996} are often used to compute mode eigenfunctions. Inhomogeneities and 3D structure in the Sun, such as flows and magnetic fields, cause the solar eigenfunctions to be different from those of the reference set. However, the reference eigenbasis has properties of completeness and orthogonality, allowing us to write solar eigenfunctions as weighted linear combinations of the reference {set. Thus}, the normal modes of the Sun are said to be ``coupled" with respect to the reference, and may be used to probe non-axisymmetric properties of the solar interior  \citep{woodard_1989}. 

Over the past fifty years, a number of helioseismic techniques have been developed to image the subsurface. Some of the most successful results include inferences of the Sun's radial and latitudinal differential rotation \citep{schou_etal_1998} using global-mode analysis, sectoral Rossby waves \citep{loeptien_etal_2018} through ring-diagram analysis \citep{hill_etal_1996}, and meridional circulation \citep{gizon_etal_2020} using time-distance helioseismology \citep{duvall_etal_1993_TD}. Each of these techniques tends to utilize one component of the wavefield, e.g. frequency shifts or travel times, whose estimation often requires significant data processing. The technique of normal-mode coupling \citep{woodard_1989}, on the other hand, contains straightforward prescriptions for retrieving seismic measurements from observations, i.e., weighted linear sums over Fourier-domain wavefield correlations, which are then ready for making inferences of subsurface perturbations. The technique has proven successful in detecting and inferring global-scale perturbations such as Rossby waves and toroidal flows \citep[e.g.][]{hanasoge_mandal_2019,hanasoge_etal_2020}. While mode-coupling theory for local helioseismic analysis has been developed \citep{woodard_2006}, with some inferences \citep{woodard_2007}, validation and detailed comparisons to other techniques has yet to be performed. 

{The local helioseismic application of mode coupling assumes the observed wave field, over a small patch of the solar surface, is described by an orthonormal and complete basis of eigenfunctions $\bxi_\alpha(\bx)$, where $\bx = (x,y,z)$ describes the Cartesian space coordinate\footnote{$x$ is oriented along the direction of rotation, $y$ along the toward the solar north pole and $z$ pointed outwards along the local radial coordinate}. The description in Cartesian geometry is valid in the limit where the observed patch size used in the analysis is much smaller than the solar radius. Here we gather the mode characteristics in the subscript $\alpha\equiv\{n,{\bf k}\}$ which denotes the radial order $n$ and horizontal wavenumbers ${\bf k} = (k_x,k_y)$ of the mode.  As mentioned earlier, subsurface perturbations in the Sun render the solar eigenfunctions $\bxi^{\odot}_\alpha$ different from those of the reference model $\bxi_\alpha$. However, the orthonormality of the reference eigenfunctions allow us to express the former as a weighted linear sum of the latter, $\bxi^\odot_\alpha=\sum_{\alpha'}c^{\alpha'}_{\alpha}\bxi_{\alpha}$. The coupling coefficients $c_\alpha^{\alpha'}$ are dependent on the unknown subsurface structures in the sun and can be estimated from observations. } 

{Under the first-Born approximation, which assumes small perturbations, previous studies have shown that coupling coefficients $c_\alpha^{\alpha'}$ are proportional to the cross-correlation of the complex Fourier components of the observed surface wave field $\phi^\omega_\bk=\phi(\bk,\omega)$ \citep[e.g.,][]{woodard_2006,woodard_2014}. In Cartesian geometry, this relationship is expressed as
\begin{equation}\label{eq.couplingcoeff}
    \langle \phi^{\omega*}_\bk \phi^{\omega+\sigma}_{\bk+\bq}\rangle = {\mathcal H}^\omega_{|\bk||\bk+\bq|\sigma}c_\alpha^{\alpha'}
\end{equation}
where the angular brackets define the expected value, $\phi^*$ is the complex conjugate of $\phi$, and $(\omega, \bk)$ and ($\sigma, \bq)$ refer to the angular frequency and vector wavenumbers of the modes and perturbations, respectively. The coefficient $\mathcal{H}$ contains information on the power-spectral model {(see Eq.\ \ref{eq.Hterm} below)} and incorporates the scattering physics described by the first-Born approximation \citep[e.g.][]{hanasoge_etal_2017}. Although $\mathcal{H}$ does not explicitly contain the radial order $n$ as a sub- or superscript, the dependence on $n$ appears through the selection of the frequency window used in computing the $\mathcal{H}$.}
The observed wave field correlations, and in turn the coupling coefficients, are dependent on the spatial structure and temporal frequency of subsurface perturbation $\mathbf{U}$. The relationship between the correlations and $\mathbf{U}$ is determined through the sensitivity kernels $\mathbf{K}$, which are derived using the reference solar model, 
\begin{equation}\label{eq.wavecorr}
    \phi^{\omega *}_{\bk}\phi^{\omega+\sigma}_{\bk+\bq} = \mathcal{H}^{\omega}_{|\bk||\bk+\bq|\sigma}\int_0^{{\rm R}_\odot}\mathbf{K}_{\bk,\bq}(z)\cdot\mathbf{U}^\sigma_{\bq}(z){\rm d}z + \eta,
\end{equation}
 {where R$_\odot$ is the solar radius and $\eta$ is a noise realization}.
 The left-hand side of Eq.~\ref{eq.wavecorr} is trivially computed from the Fourier components of a small patch of tracked Doppler images, the standard data used in most local helioseismic studies. Unlike ring-diagram analysis, which is limited to computing the mean flow field $(\sigma,\bq)=(0,0)$ within the patch, mode coupling can be used to retrieve different spatial and temporal scales. Furthermore, the method does not require the complicated geometric point-to-point cross-correlation averaging schemes of time-distance helioseismology \citep[e.g.][]{duvall_etal_1993_TD,gizon_2010}. 

In this Paper, we estimate the supergranulation power spectrum using mode coupling {(i.e. computing the left-hand side of Eq.~\ref{eq.wavecorr})} and compare the results {from} other techniques. 
Supergranulation is a distinct feature seen in the Sun's near-surface flow field and is {regarded} as an intermediate scale of thermal convection. {Supergranules} are approximately 35~Mm ($|\bq|{\rm R}_\odot\sim120$) in size, live for 1-2 days and have a strong horizontal flow component at the surface (300-400~m/s rms) \citep[see][for a review]{rincon_rieutord_2018}. While the surface properties of supergranulation have been well characterized, there remain a number of unanswered and important questions: 
(1) why does the supergranular pattern rotate at a faster rate than the magnetic features \citep{snodgrass_ulrich_1990}, 
(2) what is the depth structure of supergranules \cite[e.g.][]{duvall_hanasoge_2013,lord_etal_2014}, and
(3) what is the origin of this preferred flow scale?

Our understanding of supergranules is {rendered more complex} by the results of \citet{gizon_etal_2003}, who showed that the supergranular pattern exhibits wave-like behaviour, commonly referred to as supergranular waves. The source of these waves is yet to be understood, though their power spectrum is well observed in time-distance helioseismology \citep{gizon_etal_2003,Langfellner_etal_2018}, surface-Doppler \citep{schou_2003} and local correlation tracking of granules \citep{Langfellner_etal_2018}. Here, we will show the local-helioseismic capabilities of mode coupling and compare supergranulation spectra with those of the {earlier} studies.

\section{Data analysis}
We use seven Carrington rotations (2197-2203) of Dopplergram images taken by the Helioseismic and Magnetic Imager \citep[HMI,][]{schou_etal_2012a}. We tracked regions of size $\sim194$x194~Mm$^2$ along the equator for 11 days, from $70^\circ$ East to $70^\circ$ West. For each Carrington rotation, we generate two data cubes where the central meridian crossing time for each cube coincides with the Carrington longitude of disk center {located at} $90^\circ$ and $270^\circ$. In total, we have 14 data cubes. The spatial resolution is downsampled by a factor of 2 and the images are Postel projected. The Dopplergrams were tracked at the \citet{snodgrass_1984} rotation rate, which is 0.02893~$\mu$rad/s slower than Carrington rotation rate. In practice, the tracking and generation of data cubes is performed using \texttt{mtrack}\footnote{http://hmi.stanford.edu/rings/modules/mtrack.html}. To avoid spatial aliasing, we apply a 2D spatial-apodization function to the cubes, which is equal to 1 for pixels within 90~Mm of the patch center and tapers to zero over a distance of 7~Mm. The data cubes are then Fourier transformed in each dimension using a Discrete Fourier Transform to generate $\phi^\omega_\bk$. 


Here, we compute \Bcoeff{} coefficients which are linear-least-squares fits to raw wavefield correlations \citep{woodard_2016},
\begin{equation}\label{eq.bcoeff}
    B^\sigma_{\bk\bq} = \frac{\sum_{\omega} {\mathcal H}^{\omega*}_{kk'\sigma}\phi^{\omega*}_\bk\,\phi^{\omega+\sigma}_{\bk+\bq}}{\sum_{\omega} |{\mathcal H}^{\omega}_{kk'\sigma}|^2},
\end{equation}
where $k= |\bk|$ and $k'= |\bk'| = |\bk + \bq|$. The \Bcoeff{} coefficients are equivalent to the integral of Eq.~\ref{eq.wavecorr}. In the above equation (and Eq.~\ref{eq.couplingcoeff} and \ref{eq.wavecorr}), $\mathcal H$ is the power-spectral model
\begin{equation}\label{eq.Hterm}
    {\mathcal H}^\omega_{kk'\sigma} = -2\omega(N_{k} |R^{\omega}_{\alpha}|^2\, R^{\omega + \sigma}_{\alpha'} + N_{k'} |R^{\omega+\sigma}_{\alpha'}|^2\, R^{\omega*}_{\alpha} ),
\end{equation}
where $N_k$ is the mode amplitude and $R^\omega_\alpha$ is the Lorentzian profile of the mode resonance \citep{anderson_etal_1990,duvall_etal_1993},
\begin{equation}\label{eq.lorenz}
    R^\omega_\alpha = \frac{1}{ (\omega_{\alpha} - i\Gamma_{\alpha}/2)^2 - \omega^2 }.
\end{equation}
Here, $\Gamma_{\alpha}$ is the full width at half maximum of the mode and $\omega_{\alpha}$ is the resonant frequency. The form of Eq.~\ref{eq.Hterm} assumes the wave operator is Hermitian \citep[][]{woodard_2014}, and for simplicity, we have neglected leakage terms. The reference mode width and frequency are taken from the mode-fit parameter outputs of the ring-diagram module \texttt{rdfitc}\footnote{http://hmi.stanford.edu/teams/rings/modules/rdfitc/v13.html} \citep{basu_etal_1999}. We take fits from all tiles located at disk center from 2010 to 2018, and compute the average. The mode amplitude $N_k$ is computed through
\begin{equation}
    N_k = \frac{\sum\limits_\omega |\phi^\omega_\bk|^2}{\sum\limits_\omega |R^\omega_\alpha|^2},
\end{equation}
where the sum in $\omega$ is over the frequency bins within five line widths of $\omega_{nk}$. 

The computation of all possible \Bcoeff{} would populate a five dimensional array $[k_x,k_y,k_x+q_x,k_y+q_y,\sigma]$ for each possible $n, n'$ combination. This requires significant computational and storage capacity. However, most of these couplings would be uninformative due either to the prominence of the background or because they are insensitive to the spatial scales of interest. 
Furthermore, due to Eq.~\ref{eq.lorenz}, the coupling between modes decreases as the frequency spacing of the modes increases and the background begins to compromise the measurement. 
To ensure that we capture significant couplings, we compute \Bcoeff{} coefficients for modes sufficiently close to each other in frequency and perform the sums in Eq.~\ref{eq.bcoeff} over a number linewidths. Specifically,
\begin{equation}\label{eq.couplingLimits}
      |\omega_{\alpha} - \omega_{\alpha'}-\sigma| \leq 2\epsilon\Gamma_{\alpha},
\end{equation}
and
\begin{align}\label{eq.couplingLimits2}
    \omega \in& \{ \omega_{\alpha} - \epsilon\Gamma_{\alpha}/2,\omega_{\alpha} + \epsilon\Gamma_{\alpha}/2\} \cup\\
    &\quad\{ \omega_{\alpha'} - \epsilon\Gamma_{\alpha'}/2,\omega_{\alpha'} + \epsilon\Gamma_{\alpha'}/2\},\nonumber
\end{align}
where $\epsilon$ depends upon the $n$ and $n'$ used. Here, $\epsilon$ is determined empirically to obtain reasonable signal-to-noise ratio. {As a future study, it is useful to perform parameter searches to determine the ideal $\epsilon$.} Finally, we limit ourselves to intermediate frequencies $2\le\omega/2\pi\le4$~mHz, well below the acoustic cut-off ($\sim5.3$~mHz) and above the granulation noise ($\le2$~mHz).

\begin{figure}
    \centering
    \includegraphics[width=\linewidth]{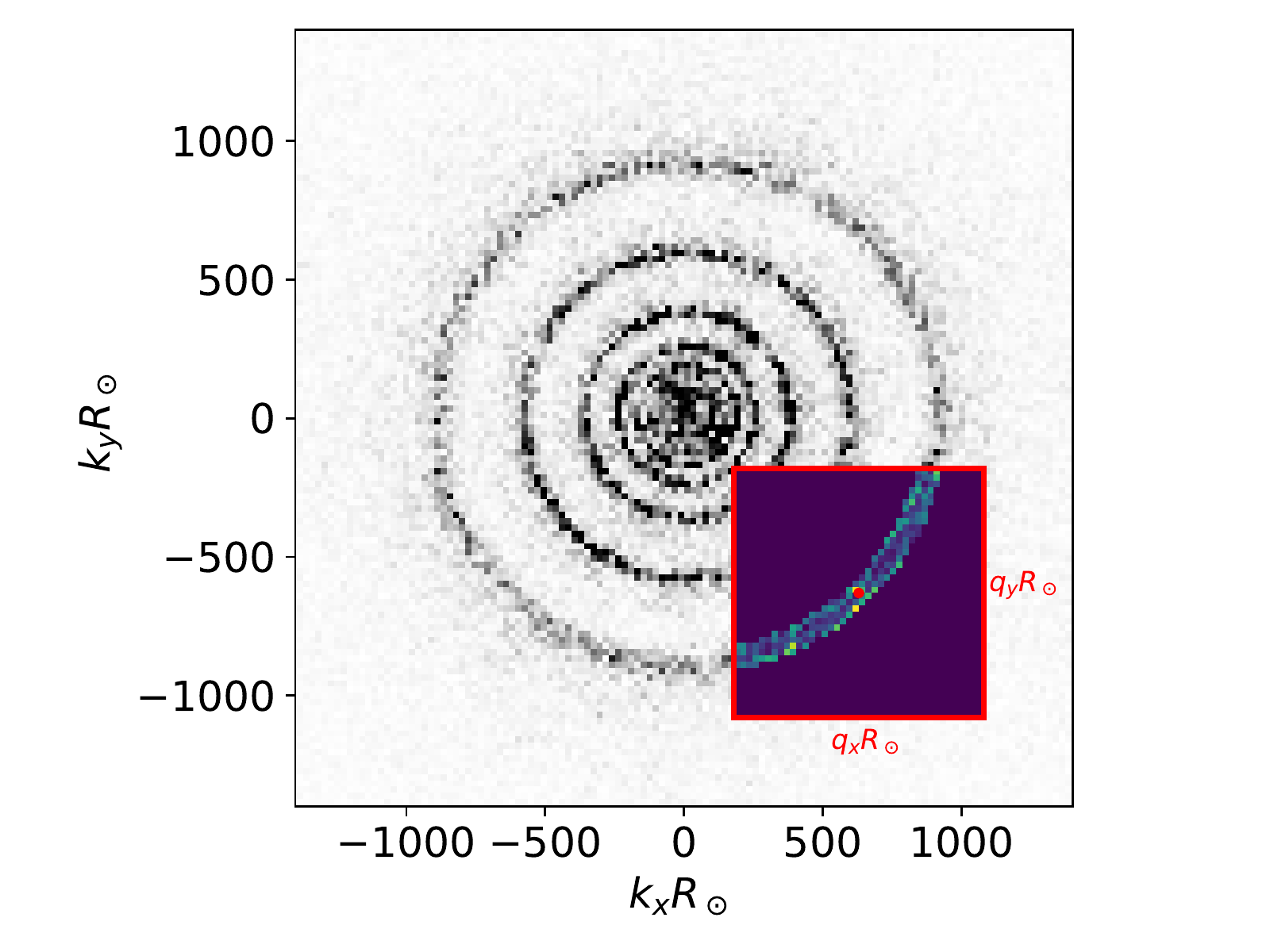}
    \caption{Slice of Doppler power spectrum at fixed frequency $\omega/2\pi\approx3$~mHz (grey scale). Inset frame shows $|B^\sigma_{\bk\bq}|^2$ computed using the couplings of an f mode $[k_xR_\odot,k_yR_\odot]=[634,-634]$ (red dot) and all other nearby f modes that meet the criteria of Eqs.~\ref{eq.couplingLimits} and \ref{eq.couplingLimits2}, where $\epsilon=1$. }
    \label{fig.coupling_concept}
\end{figure}

Figure~\ref{fig.coupling_concept} shows the power spectrum for one data cube at $\omega/2\pi=3$~mHz, as well as the squared modulus of the $B$ coefficients computed for an f mode of wave number $|k|R_\odot=900$. Given the frequency limits of Eq.~\ref{eq.couplingLimits} and \ref{eq.couplingLimits2}, the computed couplings follow the f mode ring. In this study, we estimate couplings in the range of $|\bq|\le437$, which encompasses the peak of supergranulation power $(|\bq|R_\odot\approx120)$. {Though not required for this study, one may obtain the horizontal resolution limit (half of the wavelength) by coupling opposite sides of the ring.}

\section{Noise Model}
The measurement \Bcoeff{} contains a non-trivial systematic correlated-noise component \citep[e.g.,][]{gizon_2004,woodard_2007}, which may be estimated through
\begin{equation}\label{eq.noise1}
    \begin{split}
    \langle|B^\sigma_{\bk\bq}|^2\rangle &= \frac{\sum_{\omega,\omega'} {\mathcal H}^{\omega*}_{kk'\sigma}\,{\mathcal H}^{\omega'}_{kk{'}\sigma}\langle\phi^{\omega*}_\bk\,\phi^{\omega+\sigma}_{\bk'} \phi^{\omega'}_{\boldsymbol k}\,\phi^{\omega'+\sigma*}_{\boldsymbol k{'}}\rangle}{\left(\sum_{\omega} |{\mathcal H}^{\omega}_{kk'\sigma}|^2\right)^2}.
    \end{split}
\end{equation}
Invoking Isserlis's theorem \citep{isserlis_1918}, the expected value
, denoted by angle brackets in Eq.~\ref{eq.noise1}, 
is given by
\begin{equation}\label{eq.isserlis}
    \begin{split}
    \langle\phi^{\omega*}_\bk\,\phi^{\omega+\sigma}_{\bk'} \phi^{\omega'}_{\boldsymbol k}\,\phi^{\omega'+\sigma*}_{\boldsymbol k{'}}\rangle&=\langle\phi^{\omega*}_\bk\,\phi^{\omega'}_{\bk}\rangle \langle\phi^{\omega+\sigma}_{\bk'}\,\phi^{\omega'+\sigma*}_{\bk{'}}\rangle\\ 
    &\quad\quad+  \langle\phi^{\omega*}_\bk\,\phi^{\omega+\sigma}_{\bk'}\rangle \langle\phi^{\omega'}_{\bk}\,\phi^{\omega'+\sigma*}_{\bk{'}}\rangle\\ &\quad\quad+\langle\phi^{\omega*}_\bk\,\phi^{\omega'+\sigma*}_{\bk'}\rangle \langle\phi^{\omega'}_{\bk}\,\phi^{\omega+\sigma}_{\bk{'}}\rangle.
    \end{split}
\end{equation}
We assume multivariate Gaussian noise and hence the modes are uncorrelated across different frequencies $\omega$ and wavenumbers $\bk$, i.e., $\langle\phi^{\omega*}_{\bk} \phi^{\omega'}_{\bk'}\rangle = \langle\phi^{\omega*}_{\bk} \phi^{\omega}_{\bk}\rangle\delta_{\omega\omega'}\delta_{\bk\bk'}$ \citep{gizon_birch_2004}. 
Furthermore, we neglect terms in Eq.~\ref{eq.isserlis} that contribute when $\sigma = 0$, since we are interested in finite $\sigma$.
Therefore, the noise model is expressed as
\begin{equation}\label{eq.noise}
    \langle|B^\sigma_{\bk\bq}|^2\rangle =\frac{\sum_{\omega} |{\mathcal H}^{\omega}_{kk'\sigma}|^2\,\langle|\phi^{\omega}_\bk|^2\rangle \langle|\phi^{\omega+\sigma}_{\bk'}|^2\rangle}{\left(\sum_{\omega} |{\mathcal H}^{\omega}_{kk'\sigma}|^2\right)^2},
\end{equation}
where the mode power spectrum is given by $\langle|\phi^{\omega+\sigma}_{\bk'}|^2\rangle = N_{k'}|R^{\omega+\sigma}_{k'}|^2$ and $\langle|\phi^{\omega*}_{\bk}|^2\rangle = N_{k}|R^{\omega}_{k}|^2$.


\section{Analyzing supergranulation power spectra}

\begin{table}
\centering
\caption{Mode coupling parameters}
\begin{tabular}{cccc}\hline
$n$-$n'$ coupling & $|\bk|$R$_\odot$ range & $\omega/2\pi$ range [mHz]& $\epsilon$\\\hline\hline
f-f     & 400-1000 & 2.0-3.1 &2 \\
p$_1$-p$_1$ & 400-1000 &2.5-3.9 &3 \\
p$_2$-p$_3$ & 192-750 & 2.4-4.0 &2\\\hline
\end{tabular}
\label{tab.couplingdata}
\end{table}

We compute {power spectra of the $B^{\sigma}_{\bk\bq}$ coeffcients} using $n\leq 3$ modes. {Within the spatial and temporal scales of interest, the computed power will have contributions from supergranulation, background noise and possibly yet unidentified structures.} {In this Paper we compute the spectra of f-f, \pone{}-\pone{} and \ptwo{}-\pthree{} couplings. The first two were chosen in order to compare to previous studies, while the cross $n$ pair will demonstrate that supergranulation is also encoded in correlations of these modes.} Table~\ref{tab.couplingdata} shows the range of $|\bk|$ and $\omega$ and the widths of the frequency windows {used to compute the $B^{\sigma}_{\bk\bq}$ coeffcients}. The power spectrum $P(\sigma,\bq)$ for each data cube is 
computed from Eq.~\ref{eq.bcoeff}
through
\begin{eqnarray}\label{eq.power}
        P(\sigma,\bq) = \left|\sum_{\bk} B^{\sigma}_{\bk\bq} \right|^2.
\end{eqnarray}
In this equation, we attribute equal weight to all $\bk$ when computing $P$. {Because all mode eigenfunctions are maximally sensitive to the surface, this simple averaging scheme will result in the observational sensitivity peaking near the surface.}

\begin{figure*}
    \centering
    \includegraphics[width=0.9\linewidth]{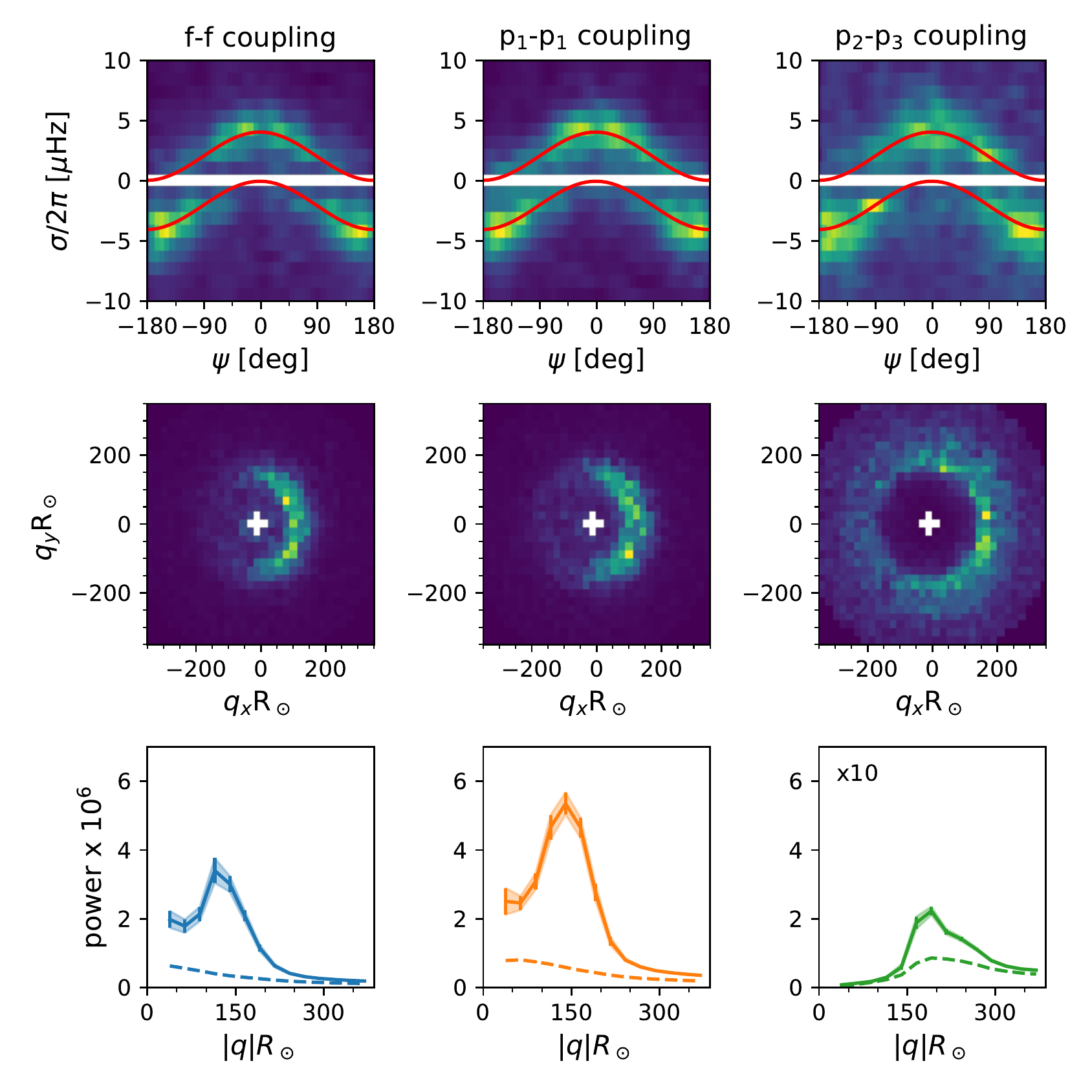}\\
    \caption{From left to right are the results from f-f, p$_1$-p$_1$ and p$_2$-p$_3$ mode coupling, respectively. Top row: Slice of power spectrum at $|q|$R$_\odot \approx172$ as a function of frequency and azimuthal distance $\psi$ from the positive $q_x$ axis. Overlaid is the dispersion relation measured by \citet{Langfellner_etal_2018} (red line). Middle row: Slice of supergranular power spectrum at $\sigma/2\pi=3.157\mu$Hz. Bottom row: average power of supergranular spectrum (solid lines) and noise model (dashed lines). Three-sigma errors on the mean are indicated by the shaded regions.}
    \label{fig.Bcoeff_power_qxqy}
\end{figure*}

Figure~\ref{fig.Bcoeff_power_qxqy} shows slices of $P(\sigma,\bq)$, averaged over the 14 data cubes, computed from three different mode-coupling pairs. Cuts of the spectra at $|\bq|$R$_\odot\approx172$ (top panels), as a function of $\sigma$ and angular distance $\psi$ from the positive $q_x$ axis, show the so-called supergranular waves. These dual peaks in frequency cannot coalesce under a Galilean transform and have thus been interpreted as traveling waves \citep{gizon_etal_2003}. Each of the mode-coupling spectra are consistent with each other and follow the measured dispersion relation of \citet{Langfellner_etal_2018}. At low temporal and spatial frequencies, especially $\sigma=0$, which we neglect, there is a systematic correlation noise that is equal to or greater than the signal.

Slices of the spectra at $\sigma/2\pi=3.157\mu$Hz (middle panels of Fig.~\ref{fig.Bcoeff_power_qxqy}) indicate that the f-f spectrum is consistent with the time-distance and LCT measurements of \citet{Langfellner_etal_2018} that show enhanced power in the prograde direction and peak power at $|\bq|$R$_\odot\approx120$. The \pone-\pone{} spectrum is similar to the f-f spectrum, though with a broader (approximately 20\%) ring of power. The \ptwo-\pthree{} spectrum differs considerably from the f-f and \pone-\pone{} spectra, with a significant depression in power for $q$R$_\odot\leq150$. This power depression is a result of $n$-$n'$ coupling, where the smallest $q$ measurable (largest flow scale) is limited by the difference in $k$ between \ptwo{} and \pthree{} at similar frequencies. For example, at $\omega/2\pi=3$~mHz, wavenumbers associated with \ptwo{} and \pthree{} differ by $k$R$_\odot=126$, and hence those couplings are less sensitive to larger scale flows ($|\bq|$R$_\odot \le 126$). {This behavior unfortunately means that some cross-$n$ couplings are inappropriate for supergranulation studies (e.g. f-\pone{} and \pone{}-\ptwo{} which are insensitive at 3~mHz to $|\bq|{\rm R}_\odot\le$302 and $|\bq|{\rm R}_\odot\le$216, respectively).} 

Frequency averages of the observed spectra (-7$\mu$Hz$\leq\sigma/2\pi\leq7\mu$Hz) and the noise models are also shown in Fig.~\ref{fig.Bcoeff_power_qxqy} (bottom panels). These averages exclude $|\sigma|\leq1.1\mu$Hz which are dominated by systematic correlated noise. The results show that the f-f couplings peak at $q$R$_\odot=120$ and become consistent with the noise model at $q$R$_\odot\approx250$. The \pone-\pone{} couplings show a 60\% greater amplitude than the f-f couplings. The peak in the \pone-\pone{} spectrum occurs at $q$R$_\odot=145$ and is consistent with the noise model at $q$R$_\odot=300$. Finally, the \ptwo-\pthree{} couplings peak at $q$R$_\odot=190$, are consistent with noise at $q$R$_\odot=300$, and have amplitudes (and noise) ten times smaller than the f-f spectrum. 
\begin{figure}
    \centering
    \includegraphics[width=\linewidth]{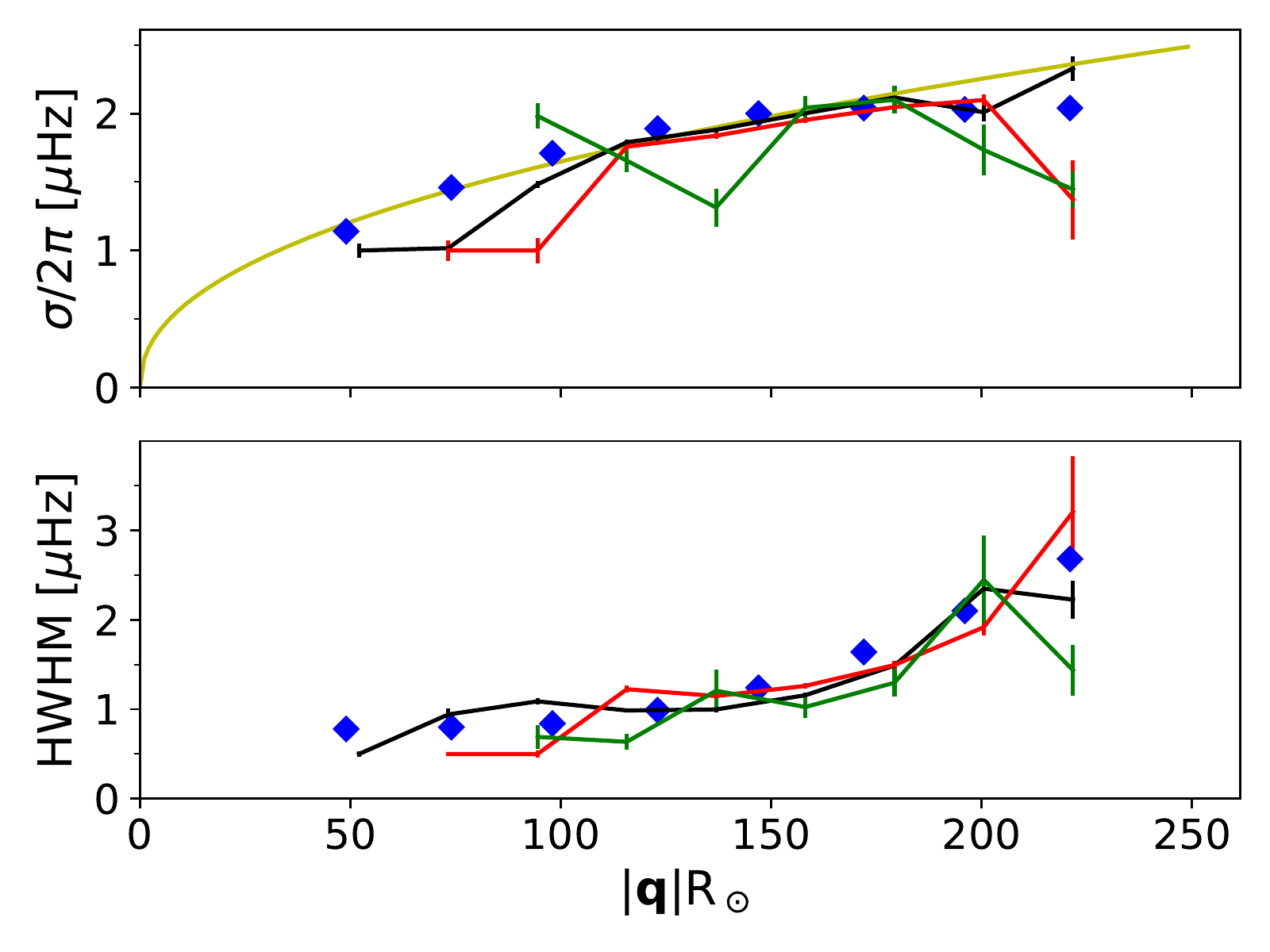}
    \caption{Dispersion relation (top) and half-width at half maximum (HWHM, bottom) obtained from Lorentzian fits of the mode-coupling supergranulation spectra. Fit parameters for the f-f  spectra (black), \pone{}-\pone{} (red) and \ptwo{}-\pthree{} (green) spectra are shown.  For reference, we show the dispersion relation of \citet{gizon_etal_2003} (yellow curve) and the fits of \citet{Langfellner_etal_2018} (blue diamonds), both computed from time-distance f-mode measurements. One-sigma errors are shown for the coupling-spectra fits. }
    \label{fig.fits}
\end{figure}

To further analyze these results we fit the spectra, for different $|\bq|$, using the parametric wave model of \citet{Langfellner_etal_2018} (see their Eqs.~1 and 2). In the coupling spectra, there is correlated noise at small $\sigma$ and $|\bq|$ (e.g. mean flow fields), which introduces a bias in the fits. To reduce this bias and improve the stability of the fitting, we increase the frequency resolution of the Doppler cubes by a factor of $\sim2.5$. This is achieved by taking each pair of data cubes in a Carrington rotation and stacking them in time, with seven days of zero padding between the cubes. This reduces the number of cubes in the averaged power spectra to seven, but confines the noise to smaller bins near zero frequency, thus improving the fit. Figure~\ref{fig.fits} shows the dispersion relation of the so-called supergranular waves, and the corresponding half-width at half maximum (HWHM, 1/lifetime). The fits from all three coupling spectra are consistent in the range of $100\le|\bq|$R$_\odot\le200$. Below $|\bq|$R$_\odot\le100$ the fits begin to be biased by the low frequency noise. Below $|\bq|$R$_\odot = 100$ the \ptwo{}-\pthree{} fits become unstable, while the \pone{}-\pone{} fits are unstable below $|\bq|$R$_\odot = 75$. Above $|\bq|$R$_\odot = 225$, the Lorentzian profiles become indistinguishable from diffuse power, and hence we do not fit beyond this scale. The measured HWHM of all three spectra remain consistent with each other for all $|\bq|$R$_\odot$. Our measured dispersion relation and HWHM show general agreement with the dispersion relation of \citet{gizon_etal_2003} ($\sigma/2\pi=1.65(|\bq|$R$_\odot/100)^{0.45}$~$\mu$Hz) and the fits of \citet{Langfellner_etal_2018}, both of which are computed from time-distance f mode measurements.




\section{Conclusions}

We have demonstrated the {straightforward nature} of normal-mode coupling  for local helioseismic studies, {and used it for detecting supergranules}. The input data for these calculations are the 3D Fourier-transformed tracked Doppler images, which also form the standard input for other local helioseismic techniques. We have shown that supergranulation power spectra can be obtained by computing the cross-correlations $\phi^*(\omega,\bk)\phi(\omega',\bk')$ of the wave field. We examined f and \pone{} self-couplings, finding that the resulting spectra are consistent with previous results \citep{gizon_etal_2003,langfellner_thesis_2015}. We also show that supergranulation may be measured in cross couplings between \ptwo{} and \pthree{}, which are capable of probing greater depths than the f or \pone{} modes. 

In all three coupling pairs, we identified the supergranular waves of \citet{gizon_etal_2003}, measuring the distinct dispersion relation that is in general agreement with that of previous studies. Our measurements of the lifetimes of these waves are also in agreement with previous findings. The nature and origin of these supergranular waves is yet to be understood, but the results of this Paper show that there is information in various p mode coupling that could be used to shed further light on this phenomenon. The agreement between our measurements of the complex spatio-temporal dynamics of supergranules and those of prior studies implies that mode coupling will be an effective technique with which to probe new and exciting questions.

While we show the spectra for only a few $n$-$n'$ pairs, there is considerably more information to exploit. The mode-parameter fitting routine of the ring-diagram pipeline (whose output is used in Eq.~\ref{eq.Hterm}) often fits up to p$_5$ {(for $15^\circ$ tiles)}, enabling the use of 21 self- and cross-coupling pairs. The different depths and horizontal scales to which these couplings are sensitive provide information on the structure of supergranulation. {Future studies that attempt to invert for the depth structure of supergranulation will need to compute all of these pairs.}{ Furthermore, the information contained in these couplings could also be used to measure the Lorentz stresses \citep[e.g.][]{hanasoge_2017} associated with supergranulation.}

The supergranular power spectrum shown here has contributions from both the toroidal and polodial components of the flow. \citet{Langfellner_etal_2015} showed that the contributions of both these components are significant. Separation of these flow components in the \Bcoeff{} coefficients requires careful analysis of sensitivity kernels for coupling \citep[derived in][]{woodard_2006} and inversions at the surface. This, and the yet-unknown depth profile of supergranules, may be addressed using mode-coupling. 

\hfill\break

The authors downloaded HMI data from the German Data Center at the Max Planck Institute for Solar System Research.
The Center for Space Science at NYU Abu Dhabi is funded by NYUAD Institute Grant G1502.
The HMI data are courtesy of NASA/SDO and the HMI Science Team. 
Processing of HMI data was performed on the DALMA compute cluster at NYUAD.

\bibliography{References}{}
\bibliographystyle{aasjournal}



\end{document}